\documentclass[12pt,preprint]{aastex}

\def\BE{\begin{equation}}
\def\BEL#1{\begin{equation}\label{#1}}
\def\EE{\end{equation}}
\newcommand{\IRAS}{{\it IRAS}}
\newcommand{\COBE}{{\it COBE}}
\newcommand{\MAP}{{\it MAP}}

\newcommand{\LPH}{LPH~201.663+1.643}

\newcommand{\HII}{H\,{\scriptsize II}}
\newcommand{\Halpha}{H$\alpha$}
\newcommand{\etal}{{\it et al.}~}

\newcommand{\degree}{^\circ}
\newcommand{\s}{{\rm ~s}}

\newcommand{\cm}{{\rm ~cm}}
\newcommand{\mm}{{\rm ~mm}}

\newcommand{\MJypSr}{{\rm ~MJy}/{\rm sr}}
\newcommand{\MHz}{{\rm ~MHz}}
\newcommand{\GHz}{{\rm ~GHz}}
\newcommand{\Hz}{{\rm ~Hz}}
\newcommand{\K}{{\rm ~K}}
\newcommand{\mK}{\rm ~mK}
\newcommand{\microK}{\mu{\rm K}}

\begin{document}

\title{Tentative Detection of Electric Dipole Emission \\ from Rapidly
Rotating Dust Grains}

\author{Douglas P. Finkbeiner\footnote{and University of California at Berkeley, Department of Astronomy}
\footnote{Hubble Fellow},
David J. Schlegel}
\affil{Princeton University, Department of Astrophysics,
Peyton Hall, Princeton, NJ 08544}
\and
\author{Curtis Frank}
\affil{University of Maryland, Department of Astronomy, College Park,
MD 20742-2421}
\and
\author{Carl Heiles}
\affil{University of California at Berkeley, Department of
Astronomy \\ 601 Campbell Hall, Berkeley, CA 94720}


\begin{abstract}
We present the first tentative detection of spinning dust emission 
from specific astronomical sources.  All other detections in the
current literature are statistical. 
The Green Bank 140 foot telescope was used to observe 10 dust clouds
at 5, 8, and $10\GHz$.  In some cases, the observed emission was
consistent with the negative spectral slope expected for free-free
emission (thermal bremsstrahlung), but in two cases it was not.
One \HII\ region (\LPH) yields a rising spectrum, inconsistent with 
free-free or synchrotron emission at the $\sim 10\sigma$ level.  
One dark cloud (L~1622) has a similar spectrum with lower
significance.  Both spectra are consistent with electric dipole
emission from rapidly rotating dust grains (``spinning dust''), as
predicted by Draine \& Lazarian.

\emph{Subject headings: }
cosmic microwave background --- 
diffuse radiation ---
dust, extinction --- 
ISM: clouds --- 
radiation mechanisms: thermal --- 
radio continuum: ISM 
\end{abstract}

\section{INTRODUCTION}

\label{sec_intro}

In the last decade, the \COBE\ satellite and several ground- and
balloon-based experiments (QMAP, Saskatoon, MAXIMA, BOOMERanG, TOCO,
DASI, CBI, and others) have greatly increased our knowledge of the
Cosmic Microwave Background (CMB) radiation.  Some of these careful
observations of the microwave sky have also revealed new and
surprising features in the interstellar medium at $14 < \nu <
53\GHz$.  
At frequencies above $100\GHz$ ($\lambda < 3\mm$), the emission from
Galactic cirrus is consistent with thermal emission.  This
emission is a broken power law, with the break at $\sim 500\GHz$,
interpreted by Finkbeiner \etal\ (1999) as emission from two dust
components.  Though this interpretation fits the data from 100 to
$3000\GHz$, a dramatic deviation arises at lower frequencies.  This
deviation motivates the work presented here.

The \COBE/DMR data (7$\degree$ FWHM) exhibit dust-correlated emission
at 90 GHz at approximately the level predicted by a detailed model of
the thermal dust spectrum based on DIRBE and FIRAS data
\cite{finkbeiner99}.  In the other DMR channels there is a pronounced
excess -- more than a factor of 10 at 31.5 GHz \cite{kogut96}.  The 19
GHz data from Cottingham's thesis \cite{cott87,boughn92} also indicate
such an excess \cite{doc98}.  This microwave excess appears in the
Saskatoon experiment as well, but with less significance \cite{doc97}.
OVRO observations in a ring around the North Celestial Pole at 14 and
31 GHz have demonstrated a correlation of 14 GHz emission with dust at
the highest resolution ($7'$) to date, but at a level $\sim 1000$
times brighter than the expected thermal (vibrational) dust emission
\cite{leitch97}.  This excess has been called the ``mystery component''
\cite{doc01}.

The spectral shape of these early measurements is consistent with
free-free emission from ionized gas, motivating a comparison with
\Halpha\ maps.  Leitch \etal\ (1997) noted the weakness of \Halpha\
emission in the OVRO fields and concluded that only $T>10^6\K$ gas
(e.g. shock-heated gas in a supernova remnant) could produce the
observed emission.  However, the observed emission would require an
energy injection rate at least 2 orders of magnitude greater than that
provided by supernovae (Draine \& Lazarian 1998a).  Another
possibility is magnetic dipole emission\footnote{
Note that this mechanism has nothing to do with grain rotation.} 
from ferromagnetic or ferrimagnetic grains, resulting from thermal
fluctuations in the magnetization of the grain material.  This
component is certainly present at some level, but is sub-dominant
in the current data \cite{draine99}.  

The currently favored emission mechanism is electric dipole emission
from rapidly rotating dust grains, an idea first proposed by 
Erickson (1957) and improved upon by Ferrara \& Dettmar (1994).  Recent
work by Draine \& Lazarian (1998b) has refined this idea and shown
that ion encounters with dust grains are the dominant spin-up
mechanism, leading to concrete predictions of emission as a
function of the temperature, density, and ionization fraction of the
surrounding gas.
This emission mechanism
has never been unambiguously observed, but can be made to agree with
the observed microwave data for reasonable model parameters.  It was previously
impossible to tell if the excess emission results from spinning dust
or free-free emission, because the predicted spectral slope is similar
for $20 \la \nu \la 40$ GHz, and earlier observations were
insufficient to rule out either mechanism.  The Tenerife data
\cite{doc99} at 10 and $15\GHz$ have good leverage on the spectral
shape, but provide only a statistical detection by cross-correlation.
To date, no one has observed spinning dust emission from a specific
source.

The spinning dust is expected to have an emission peak at $\sim
15\GHz$ (in temperature units) and be dominated by free-free below
$\sim 5-10\GHz$, so by choosing frequencies in this range, one can
hope to unambiguously detect this component.  We have obtained data
for several dust clouds at 5, 8, and $10\GHz$, a frequency range over
which the spinning dust spectrum differs measurably from that of
free-free or synchrotron emission.

%
\section{OBSERVING STRATEGY}

This study used the NRAO\footnote{
The National Radio Astronomy Observatory is a
facility of the National Science Foundation operated under cooperative
agreement by Associated Universities, Inc.}
43m (``140 foot'') telescope at Green Bank
shortly before its decomissioning on 1999 July 19.  Cassegrain C-band
($5\GHz$) and X-band ($8-10\GHz$) receivers were used with a $300\MHz$
bandwidth, and gave a typical system temperature of $30-60\K$.

\begin{figure}[tb]
\plotone{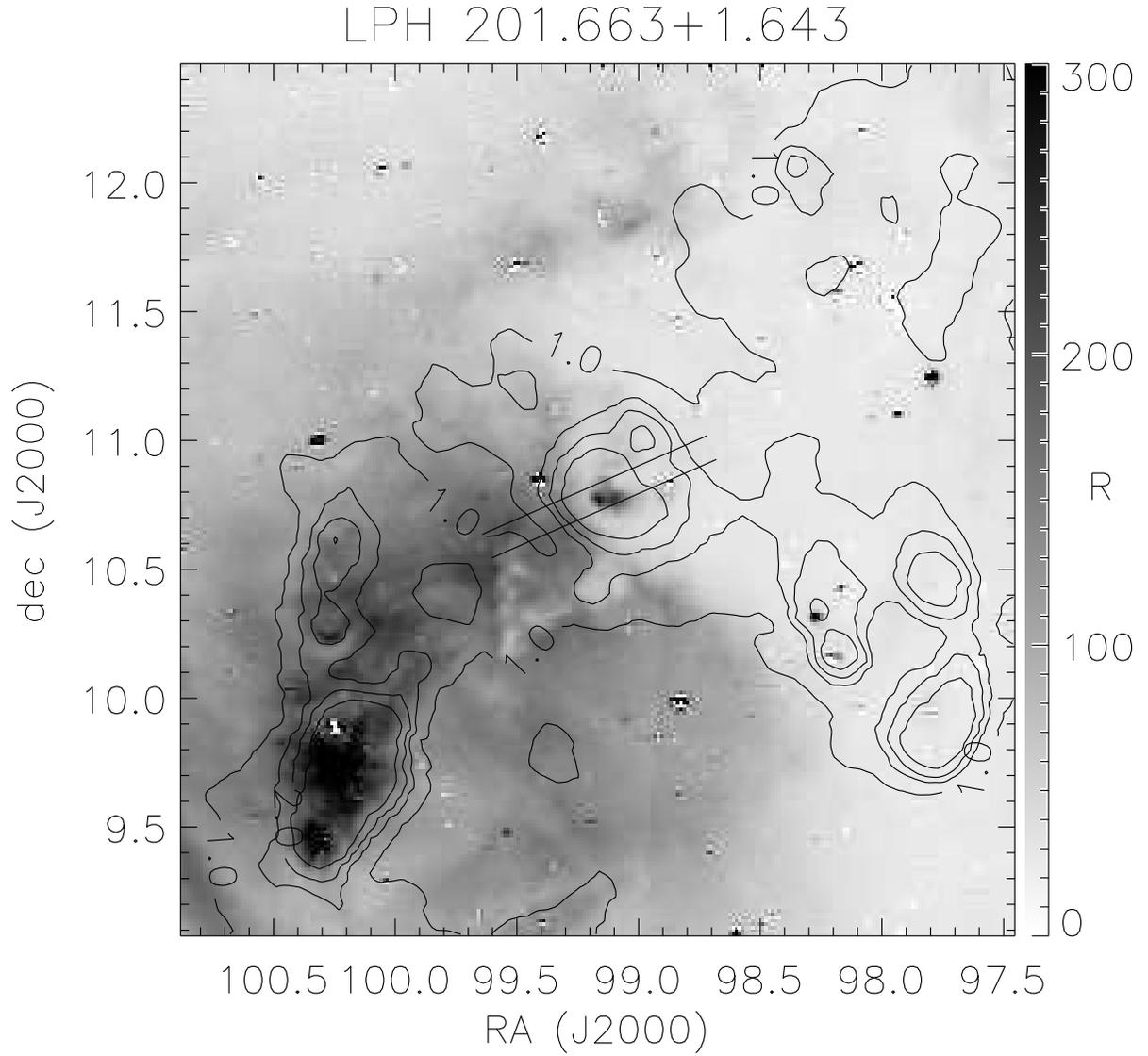}
\figcaption{Scan location for the LPH cloud (\emph{black
lines}), overplotted on SHASSA (Gaustad \etal\ 2001) \Halpha\
(\emph{grayscale}) and SFD98 $E(B-V)$
\label{fig_LPH_field}
(\emph{contours}).  Contour levels are 1,1.5,2,3 mag.  \Halpha\
brightness is measured in Rayleighs with the scale on the right.  
One Rayleigh is $10^6/4\pi$ photon
${\rm cm^{-2} s^{-1} sr^{-1}}$. }
\end{figure}

\begin{figure}[t]
\plotone{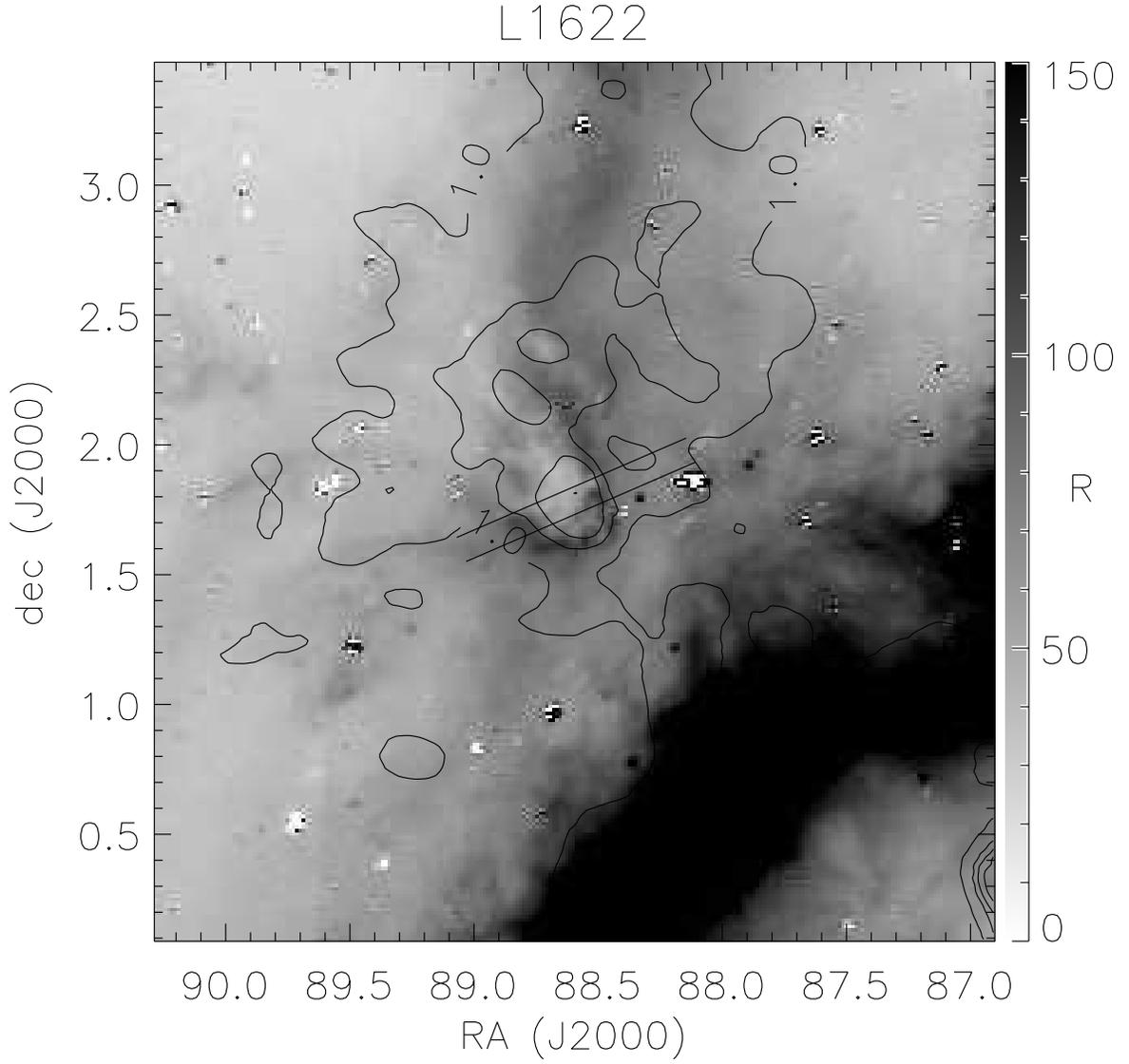}
\figcaption{
\label{fig_L1622_field}Scan location for Lynds 1622 (\emph{black lines}), overplotted on
\Halpha\ (\emph{grayscale}) and SFD98 $E(B-V)$ [mag] (\emph{contours}).  }
\end{figure}

Data were obtained during two runs: 22-27 April and 1-6 June, 1999. 
The nutating subreflector on the 140 foot can switch at up to $2.5\Hz$
with a $12'$ throw.  Switching is restricted by hardware to a position
angle of $292.5\degree$ (E of N) for the X-band ($8-10\GHz$) receiver.  A
second level of switching was accomplished by driving the telescope
along the PA$=292.5\degree$ direction at a rate of $3' \s^{-1}$,
completing a $48'$ scan every $16\s$  (Figures
\ref{fig_LPH_field} and \ref{fig_L1622_field}).
During the scan, the ON--OFF
difference was recorded by a digital continuum recorder (DCR) for
every chop.  The same strip was then observed in the ``reverse''
direction, a sequence requiring nearly 1 min for a round trip
(including turn-around time).  During the turn-around time, a calibration
noise source was flashed at 1 Hz.  In C-band ($5\GHz$), switching is
restricted to a direction orthogonal to the scan line, so switching
was not performed at $5\GHz$.



The beam of the 140 foot telescope at $5\GHz$ ($6'$ FWHM) is
well-matched to the SFD98 $100\micron$ map, to which the data are
compared in \S\ref{sec_analysis}.
At 8 and $10\GHz$,
multiple parallel scans were performed to subsample a $6'$ Gaussian
beam.  For a typical system temperature of 40K and 200ms integration
(2Hz switch, so 4 samples per second with 50ms blanking time) we
obtain a theoretical sensitivity of 7.3 mK per difference per
polarization.  Typically 1 hr was spent on each target in each band,
yielding a theoretical sensitivity of $\sim 1\mK$ per polarization per
pointing.  Observations were also attempted at $18\GHz$, but the system
temperature was too high to obtain any useful data.  The $14\GHz$
receiver was not available at the time of our observing run.

A bandwidth of $300\MHz$ was used throughout the run.  RFI affected a
negligible number of the measurements, and these are easily found and
discarded later in the analysis.

\section{ANALYSIS}
\label{sec_analysis}
The Digital Continuum Recorder (DCR) records the difference between
the ON position and OFF
position some $12'$ away as the telescope scans over a target.  The PA
of the chop is fixed at $292.5\degree$, so as not to interfere with
point source measurements when scanning E-W or N-S.  This arrangement
is ideal for point sources, because the OFF never crosses the
source and the baseline difference is easily established when the ON position
is away from the source.  When observing the diffuse ISM, there is no
local zero-brightness position to compare to along the scan.  We
therefore scan parallel to the chop direction to reduce the problem to
one dimension.

\begin{figure}[t]
\plotone{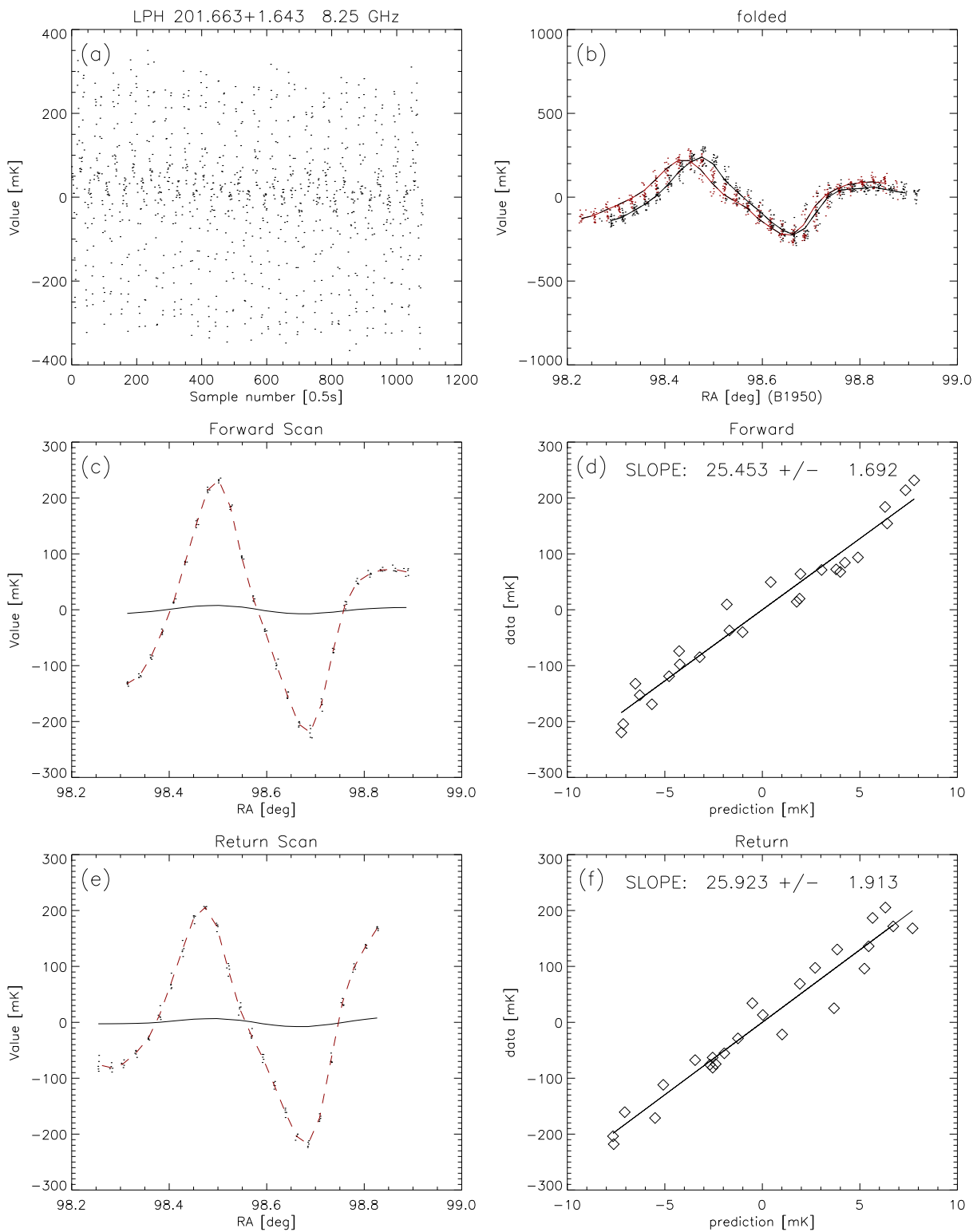}
\figcaption{
\label{fig_LPH}
For the diffuse \HII\ region \LPH: 
($a$) observed differences vs. time;
($b$) differences folded vs. RA for (\emph{black}) forward and 
(\emph{red}) return scans;
($c$) (\emph{dashed}) data and (\emph{solid}) SFD-based prediction
using conversion factor for $10\GHz$ from de Oliveira-Costa \etal\
(1999) for the forward scan;
($d$) correlation of data vs. prediction for the forward scan;
($e$) same as ($c$) for the return scan; and 
($f$) same as ($d$) for the return scan. 
Correlation slopes would be unity at $10\GHz$ for a correct
prediction, and be somewhat less at $8\GHz$.
Only Right Circular Polarization data are shown. 
LCP data look similar and there is no evidence for circular
polarization.
}
\end{figure}

The measured difference for each chop (Figure \ref{fig_LPH}$a$) is
dominated by sidelobe and atmospheric contamination, and drifts with
time.  All these effects are generally a smooth function of time, and
are removed with a low order polynomial fit with appropriate
outlier rejection.  
The data are then ``folded'' by plotting as a function of sky position
(Figure \ref{fig_LPH}$b$) to reveal the beam-convolved and differenced
structure on the sky.  Because of sidelobe and atmospheric
contamination, an artificial slope is present in these plots; it is
also fit and removed.  The data are binned in RA and the median of
each bin is overplotted.  Note that for the source
\LPH\ shown in Figure \ref{fig_LPH}, there is a difference between forward and
reverse scans.  In the forward direction, the ON position is slightly
less than 12' ahead of the OFF position because the telescope is
moving east.  In the reverse direction, ON is slightly more than 12'
away, as the telescope drives west.  This means that forward and
reverse scans must be analyzed separately, even though they are
overplotted in Figure \ref{fig_LPH}$b$.

Finally, we consider the correlation slope between predicted and
observed emission.  Because of the double-switched observing strategy,
it is impossible to uniquely recover the observed flux of an object.
However, a suitable template may be convolved with the same observing
strategy and compared to observation.  The SFD98 $100\micron$ and
temperature maps are used to predict a microwave brightness
temperature (Figures \ref{fig_LPH}$c$ and \ref{fig_LPH}$e$) using the factor of $50\microK$ at
$10\GHz$ per $I_{100\micron}$ in$\MJypSr$  \cite{doc99}.  The correlation
slope is then measured and tabulated; it should be consistent with 1
at $10\GHz$ for a correct prediction, and vary at other frequencies
with the shape of the spinning dust spectrum.  Separate numbers are
tabulated for each combination of polarization and direction
(Table \ref{table_slopes}).  The prediction includes a factor of $1/2$
for comparison to the single-polarization data.  This is why the LCP
and RCP measurements are combined by averaging rather than adding
them.  The (total Stokes I) $T_B$ per $I_{100}$ in units of $\microK
(\MJypSr)^{-1}$ may be obtained from Table \ref{table_slopes} by
multiplying the correlation slopes by 50.  Note that L1622 was
observed on two occasions at 8.25 GHz.  

\subsection{Calibration}
For calibration standards, we use the fits of Salter (2001), who follows
Kuehr \etal\ (1981) in using the form $\log_{10}S=a_0+a_1x+a_2\exp(-x)$ where
$x=\log_{10}\nu$, $\nu$ is in MHz and $S$ is in Jy. 
With $a_0=3.523$, $a_1= -0.779$, and $a_2= -3.732$ for 3C138, we
obtain S(5, 8.25, 9.75 GHz) = (3.54,2.50,2.22) Jy.  This agrees with a
simple power-law fit to 87GB \cite{gb87} and the Wright \etal\ (1991)
survey to within 2\%.  Cross-checks were done with 3C245 and BL-LAC,
and agreed to 5 percent.  The definitive calibration was determined by 
3C138 however, because it was observed near L1622 and \LPH\ in both
space and time. 

\section{RESULTS}

Of the targets observed (Table 1) two clouds show significant
dust-correlated emission at $5-10\GHz$.  The non-detection of the
other targets is not very surprising, because the parameters describing
the physical properties of these clouds span a wide range of values,
so the relative intensities may vary widely.
Both of the detected clouds (Lynds 1622 and \LPH) show a steep rise from
$8-10\GHz$, with the $5\GHz$ brightness apparently contaminated by
free-free (Table \ref{table_results}).
In order to compare these results to previous correlation
measurements, results from several experiments are overplotted on the
Draine \& Lazarian models in Figure \ref{fig_result}.
All data and model curves are normalized to emission per H
atom for ease of comparison.  In practice this is done by normalizing
to SFD $E(B-V)$ values, and then applying a conversion factor.  
The free-free emission curves corresponding to two values of $<n_e
n_p>/<n_H>$ are shown, with the upper curve determined by an emission
measure derived from the \Halpha\ survey\footnote{
The data are available at ``http://amundsen.swarthmore.edu/SHASSA/''}
of Gaustad \etal\ (2001).

Because free-free emission clearly contributes to the measured signal,
it is desireable to have an independent limit on the free-free and
show that it is consistent with our measurements.  This limit may be
derived from \Halpha as follows. 
The emission coefficient $j_\nu$ for free-free, with electrons assumed
to interact with ions of charge $Z_i e$ and particle density $n_i$ is
\BE
j_\nu = 5.44 \times 10^{-16} \frac{g_{ff}Z_i^2 n_e n_i}{T^{1/2}}
e^{-h\nu/kT} {\rm Jy~sr^{-1}~cm^{-1}}
\EE
where $g_{ff}$ is the gaunt factor for free-free.  For microwave
frequencies, a useful approximation is
\BE
g_{ff} =
\frac{3^{1/2}}{\pi}\left[\ln\frac{(2kT)^{3/2}}{\pi e^2\nu m_e^{1/2}}
 - \frac{5\gamma}{2}\right]
~~~~~\nu_p \ll \nu \ll kT/h
\EE
where $\gamma$ is the Euler constant ($\gamma\approx 0.577$) and
$\nu_p$ is the plasma frequency
(Spitzer 1978, p. 58).
Evaluating for $\nu = 10^{10}\Hz$ and $T=10^4\K$ we find
$g_{ff}\approx4.69$ and $J$($10\GHz$) = 78.8 Jy sr$^{-1}$ for EM = 1 pc
cm$^{-6}$.
The emission measure for these clouds may be estimated from the
\Halpha\ emission shown in Figures 1 and 2.  
Note that 1R = $10^6/4\pi$ photon ${\rm cm^{-2} s^{-1} sr^{-1}}$
and corresponds to EM $\approx$ 2 pc cm$^{-6}$ for \Halpha\ at $T=10^4\K$.
For \LPH, the peak of the cloud is about 200R or EM=400.  Using a 
conversion factor of N(H)=$8\times10^{21}$ for 1 mag $E(B-V)$, we get
free-free emission of $3.2\times10^4$ Jy/sr for $N(H)=2.4\times10^{22}$ or 
$j_\nu/n_H = 1.3\times10^{-18}\ {\rm Jy~sr^{-1}~atom^{-1}}$ at
$10\GHz$. 
The SFD98 map shows about 6 mag extinction ($\tau \approx 6.5$) for an
\Halpha\ photon
traveling completely through this cloud, so this free-free
estimate should be taken as a lower limit for photons from within the
cloud. 
If the \Halpha\ emitting material is uniformly mixed with dust in a
cloud with total optical depth $\tau$, the effective flux observed is 
\BE
J_{\nu,eff} = \frac{J_\nu}{\tau}\int_0^\tau d\tau'e^{-\tau'}
=\frac{J_\nu}{\tau}(1-e^{-\tau})
\EE
In the limit of large $\tau$, the flux is simply reduced by a factor
of $\tau$.  In Figure \ref{fig_result}, the level of predicted
free-free has been multiplied by a factor of 7 and appears to be in
good agreement with the uniform mixture hypothesis.  Of course, this
result is very sensitive to the configuration of material on the front
side of the cloud, so the agreement may be largely coincidental.


\section{FUTURE WORK}
\subsection{Requirements}
In order to demonstrate convincingly that the excess emission detected
by this work is electric dipole emission from spinning dust, the
following requirements are proposed.

1.  Observations for any line of sight must agree with the model over
    a wide frequency range ($5-60\GHz$) for reasonable values of the
    model parameters (gas and dust temperature, density, and
    ionization fraction).  The observations must be inconsistent with
    free-free alone at high confidence.  Simply stated, the spectral
    shape should exhibit a ``roll-off'' on both sides of the emission peak.

2.  Variation in the spectrum of spinning dust emission should trace
    variation in the parameters as derived from spectral line
    information and the SFD maps.  The nature of this variation should
    be (at least qualitatively) similar to that anticipated by the
    Draine \& Lazarian model.

3.  Alternative explanations for the excess (such as magnetic dipole
    emission) should be ruled out by the spectral shape.
    Polarimetry is also useful, as the spinning dust emission is
    relatively unpolarized (Lazarian \& Draine 2000).  

None of these requirements are met by the current paper, so our
interpretation of the observed excess is only tentative.  During the
next few years, however, dramatically improved data will become
available.

\subsection{Prospects}
There are compelling reasons to pursue this project with an
interferometer in the southern hemisphere, such as CBI (Cosmic
Background Imager, Padin \etal\ 2001).  The synthesized beam
is $5-8'$, well-matched to the $6'$ of the IRAS \cite{iras88} map as
reprocessed by Schlegel \etal (1998).
CBI has 10 channels from $26-36\GHz$, providing spectral information
with high sensitivity.  Many prospective targets are in the southern
half of the sky, toward the Galactic center.  The frequency coverage
is not ideal for separation of free-free from spinning dust, but the
high sensitivity ($41\microK$ in 900s for the highest resolution
configuration) should allow clean detections and easy comparison with
the Rhodes \cite{jonas98} survey at $2.326\GHz$ for free-free removal.
The DASI (Degree Angular Scale Interferometer, Halverson \etal\ 1998) 
has many of the same properties, but a larger beam of about $20'$, and
is therefore more appropriate for comparison with the Rhodes data
($20'$).  

The recently launched Microwave Anisotropy Probe (\emph{MAP;} Bennett \etal\
1997) has full-sky
coverage at 22, 30, 40, 60, and 90$\GHz$ with beams of 56, 41,
32, 21, and
$14'$ (FWHM) respectively.  These data will complement the interferometer data 
nicely.  With data from these projects, a decisive detection
of spinning dust emission may be possible in the near future.

\section{CONCLUSIONS}
We have explored a FIR-selected sample of dust clouds spanning a wide
range of IR color-temperature, column density, and ionization
fraction.  Our target selection procedure rejected the radio-bright
\HII\ regions as poor targets, because free-free was expected to
overwhelm spinning dust emission.  In spite of this prejudice, one
\HII\ region (from the catalog of diffuse \HII\ regions in Lockman,
Pisano \& Howard 1996; LPH) made the list, and provided a very
significant detection.  A dark cloud, Lynds 1622, was also detected
and found inconsistent with free-free emission at a lower confidence.
This is the first detection of a rising spectrum source at $8-10\GHz$
consistent with spinning dust.  The amplitude of this emission per H
atom apparently varies by a factor of at least 30, somewhat more than
theoretically expected.  Some adjustment of model parameters will be
needed to explain the brightness of \LPH, but the current model of
Draine \& Lazarian can probably accommodate the new data (Draine,
2001).  Note that magnetic dipole emission from ferromagnetic or
ferrimagnetic grains \cite{draine99} is substantially weaker than the
observed signal from \LPH, but significant contributions from such a
mechanism cannot currently be ruled out.\footnote{
Because this emission may be
$\sim30\%$ polarized for aligned grains of strongly magnetic material,
polarization measurements would be helpful.}
Other attempts have been
made to explain the large variation using $12\micron$ emission as a
tracer of the small grain population \cite{doc01}.  We caution that
the interpretation of these data must remain tentative until a larger
sample meeting the criteria in \S 5 can be obtained. 

Cold neutral clouds are an obvious target choice for future work,
because free-free
emission is likely to be subdominant in them.  Diffuse ionized gas
also appears to be an excellent target, since ion-dust interactions
are effective at spinning up the grains.  As long as the density is
low, as in the LPH list (Lockman \etal\ 1996), \HII\ regions may be
the optimal targets for future work with DASI, CBI, and \MAP.

This measurement is rather tenuous, as was the discovery of Galactic
radio emission 60 years ago \cite{reber40}.  The signal was
immediately interpreted as free-free \cite{henyey40} but was later
recognized as synchrotron radiation.  Those emission mechanisms are
now essential tools for ISM research in the Milky Way and distant
galaxies.  With the expected increase in astronomical capability at
microwave frequencies in the near future, we anticipate that spinning
dust will make a useful addition to this toolbox.

We would like to thank the staff at Green Bank, especially Dana
Balser, who spent 20 hours helping us with the backend software on the
first day.  Bruce Draine provided model emissivities and helpful
discussions.  Peter McCullough and John Gaustad provided \Halpha\ data
in the regions of interest prior to publication.  
Shaul Hanany, Al Kogut, Alex Lazarian, Jay Lockman, and George Smoot
provided helpful comments on the manuscript. 
Computers were
partially provided by a Sun AEGP Grant.  DPF is a Hubble Fellow, and
was partially supported by NASA grant NAG~5-7833.  DJS is partially
supported by the Sloan Digital Sky Survey.  This work was also
supported in part by NSF grants 95-30590 and 00-97417 to CH.  This
research made use of the NASA Astrophysics Data System.  The Green
Bank 140 foot telescope was operated by the National Radio Astronomy
Observatory (NRAO), which is a facility of the National Science
Foundation operated under cooperative agreement by Associated
Universities, Inc.

\newpage

\bibliographystyle{unsrt}
\bibliography{gsrp}


\newpage
\begin{deluxetable}{l|rrrl}
\tablewidth{0pt}
\tablecaption{Target list
   \label{table_targetlist}
}
\tablehead{
   \colhead{Name}    &
   \colhead{$\alpha_{2000}$}    &
   \colhead{$\delta_{2000}$}  &
   \colhead{result}   &  
   \colhead{comment}  
}
\startdata
%
L1622             &  05 54 23 & $ +01$ 46 54  & detected & dark cld    \\
\LPH              &  06 36 40 & $ +10$ 46 28  & detected & diffuse \HII  \\
L1591             &  06 09 55 & $ +13$ 44 34  & ND       & dark cld    \\
IRAS 07225$-1617$ &  07 24 47 & $ -16$ 23 21  & weak     & IRAS source \\
IRAS 15522$-2540$ &  15 55 16 & $ -25$ 49 40  & no 10    & IRAS source \\
VSS II-79         &  16 33 58 & $ -23$ 43 48  & neg corr & near L1709C \\
IRAS 18146$-1200$ &  18 17 29 & $ -11$ 58 52  & 8 only   & near \HII \\
IRAS 19441$+2926$ &  19 46 08 & $ +29$ 33 34  & ND       & IRAS source \\
IRAS 23339$+4811$ &  23 36 23 & $ +48$ 28 01  & ND       & IRAS source \\
IRAS 23350$+4815$ &  23 37 28 & $ +48$ 32 18  & ND       & IRAS source \\
\enddata
\tablecomments{
List of targets observed.  ND = not detected.  Most of these have
other names in the literature.  VSS II-79 has a significant negative
correlation with dust emission, perhaps from an \HII\ shell around the
dust filament.  
LPH = Lockman, Pisano, \& Howard (1996), IRAS = IRAS Science Working 
Group (1985), L = Lynds (1962), and VSS = Vrba, Strom, \& Strom (1976).
}
\end{deluxetable}

\newpage
\begin{deluxetable}{l|rrrrrrr}
\tablewidth{0pt}
\tablecaption{Correlation slopes
   \label{table_slopes}
}
\tablehead{
   \colhead{Name}    &
   \colhead{$\nu$}    &
   \colhead{For RCP}  &
   \colhead{Ret RCP}  &
   \colhead{For LCP}  &
   \colhead{Ret LCP}  &
   \colhead{Avg}  &
   \colhead{nsig}  
}
\startdata
L1622   &5.00&$   1.29 \pm 0.39$&$   3.49 \pm 0.84$&$   0.48 \pm 0.38$&$   3.05 \pm 0.77$&$     1.31 \pm 0.25$&   5.3 \\
L1622   &8.25&$   1.25 \pm 0.29$&$   0.26 \pm 0.37$&$   0.67 \pm 0.41$&$   0.52 \pm 0.35$&$     0.75 \pm 0.17$&   4.4 \\
L1622   &8.25&$   1.14 \pm 0.37$&$   1.24 \pm 0.44$&$   1.05 \pm 0.38$&$   1.11 \pm 0.33$&$     1.13 \pm 0.19$&   6.1 \\
L1622   &9.75&$   1.65 \pm 0.67$&$   0.76 \pm 0.77$&$   0.84 \pm 0.73$&$   0.78 \pm 0.65$&$     1.03 \pm 0.35$&   2.9 \\
LPH     &5.00&$  53.16 \pm 5.28$&$  57.01 \pm 5.68$&$  54.12 \pm 5.39$&$  58.05 \pm 5.90$&$    55.41 \pm 2.77$&  20.0 \\
LPH     &8.25&$  25.45 \pm 1.69$&$  25.92 \pm 1.91$&$  29.16 \pm 1.93$&$  29.50 \pm 2.25$&$    27.22 \pm 0.96$&  28.4 \\
LPH     &9.75&$  23.96 \pm 2.09$&$  23.33 \pm 1.70$&$  27.89 \pm 2.49$&$  27.17 \pm 1.90$&$    25.25 \pm 0.99$&  25.4 \\
\enddata
\tablecomments{Correlation slopes for Forward and Return scans, 
RCP and LCP polarizations.
These correlation slopes are for $T_B$ vs. a prediction of
$50\microK/I_{100}$ where $I_{100}$ is the DIRBE-temperature-corrected
IRAS intensity at $100\micron$ in $\MJypSr$.  This
temperature-corrected map may be obtained by dividing the SFD98
$E(B-V)$ prediction by 0.0184.  The prediction used includes a factor
of $1/2$ for single polarization measurements, so RCP and LCP are
combined by averaging, not adding.  Values in the table may be multiplied
by 50 to obtain units of $\microK/I_{100}$ in order to compare to
e.g. de Oliveira-Costa \etal\ (1999). Note that Lynds 1622 was
observed twice at
$8.25\GHz$.
}
\end{deluxetable}
\begin{deluxetable}{l|r|rrr|rr}
\tablewidth{0pt}
\tablecaption{Emissivity per H
   \label{table_results}
}
\tablehead{
   \colhead{Name}    &
   \colhead{$N(H)$}  &
   \colhead{$j_\nu(5\GHz)/n_{\rm H}$}    &
   \colhead{$8.25\GHz$}  &
   \colhead{$9.75\GHz$}  \\
   \colhead{} &
   \colhead{$\times10^{21}\cm^{-2}$} &
   \colhead{Jy sr$^{-1}$ H$^{-1}$} &
   \colhead{Jy sr$^{-1}$ H$^{-1}$} &
   \colhead{Jy sr$^{-1}$ H$^{-1}$} 
}
\startdata
L1622 & 18 &$ 7.8\pm 3.9\times10^{-19}$ &$ 6.6\pm 1.3\times10^{-19}$
&$ 10.1\pm 3.4\times10^{-19}$  \\
LPH   & 27 &$ 147\pm 17\times10^{-19}$ &$ 191\pm 7\times10^{-19}$ &$ 247\pm10\times10^{-19}$  \\
\enddata
\tablecomments{Emissivity.  $N(H)$ is estimated by using the 
SFD98 $E(B-V)$ extinction estimate and a conversion factor of 
$N(H) = 8\times 10^{21}\cm^{-2}$ per magnitude. 
}
\end{deluxetable}


\begin{figure}[t]
\plotone{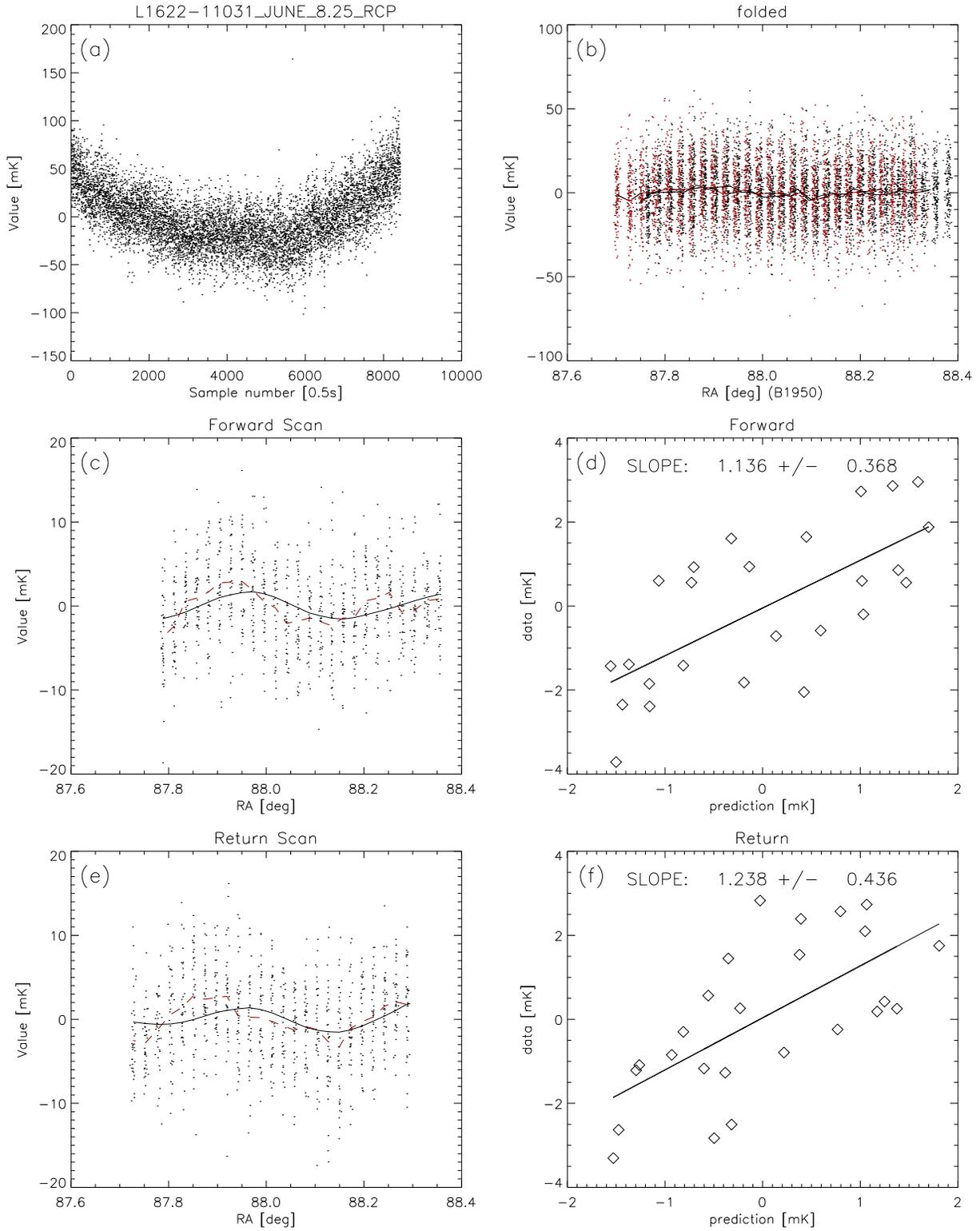}
\figcaption{
Same as Figure \ref{fig_LPH} but for Lynds 1622.  Notice that
the correlation slope is substantially less than unity in this
region.}
\label{fig_L1622}
\end{figure}


\begin{figure*}[t]
\plotone{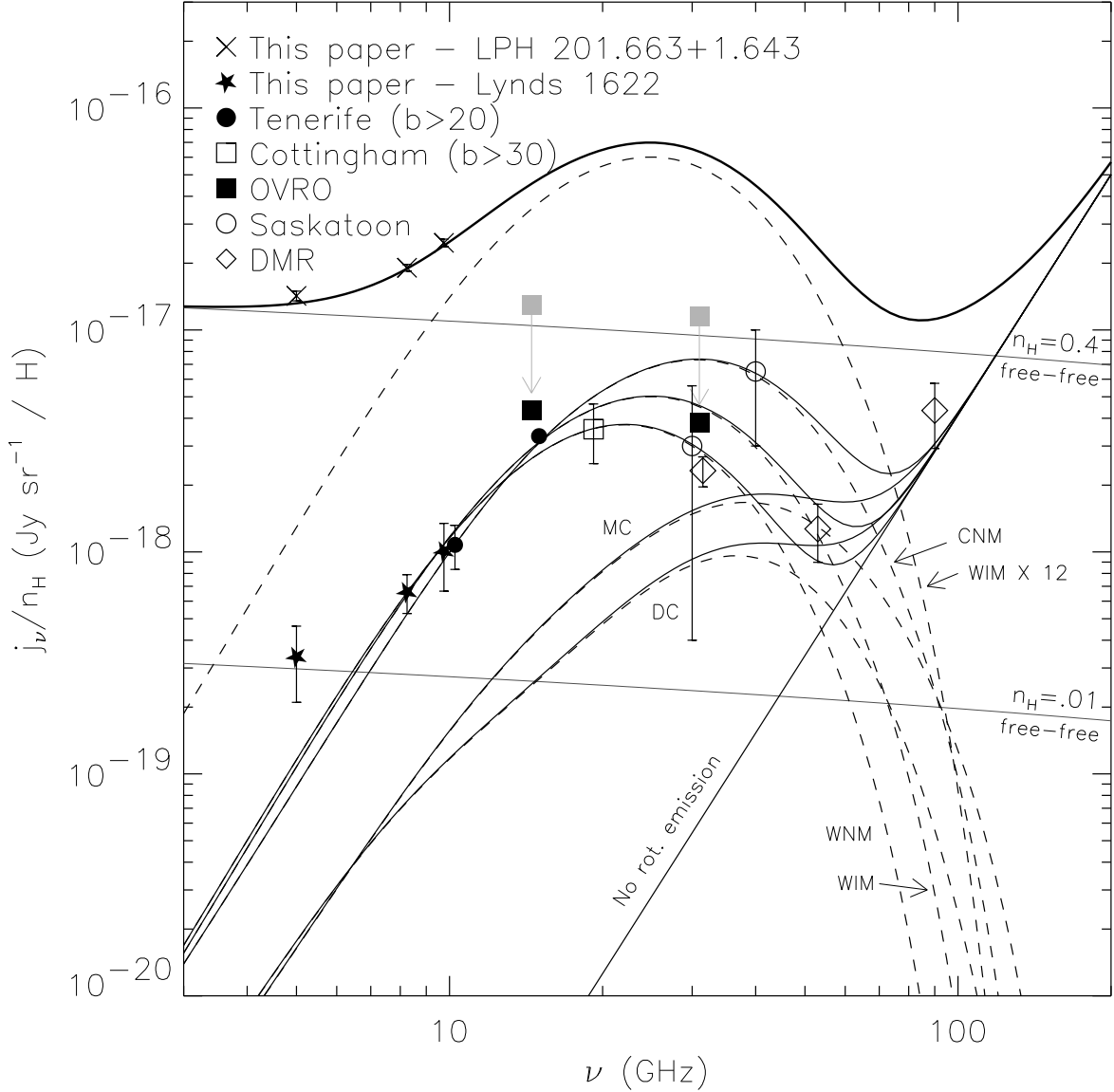}
\caption{Model Dust emissivity per H atom for DC, MC, CNM, WNM, and
WIM conditions (as in Draine \& Lazarian 1998b, Figure 9).  Solid thin
lines are total
emissivity; dashed lines are rotational emission. 
Gray lines are emission from free-free for given $n_H$, or rather 
$<n_en_p>/<n_H>$ averaged along the line of sight. 
The top thick line is the sum of free-free, vibrational dust, and 10 times
the WIM spinning dust model, shown for reference.
Also shown are measurements from the \COBE/DMR (\emph{open diamonds})
from Finkbeiner \etal\ (1999), similar to Kogut
\etal\ (1996); Saskatoon (\emph{open circles}) (de Oliveira-Costa 
\etal\ 1997); the Cottingham \& Boughn $19.2\GHz$ survey 
(\emph{open square}) (de
Oliveira-Costa \etal\ 1998), OVRO data (\emph{solid squares}) 
(Leitch \etal\ 1997);
Tenerife data (\emph{solid circles})(de Oliveira-Costa \etal\ 1999);
and this paper: \LPH\ (\emph{crosses}) and L1622 (\emph{stars}). 
The OVRO points have been lowered a factor of 3 relative to Draine \&
Lazarian (1998b, Figure 9), because the unusual dust temperature near
the NCP caused an underestimate of the H column density along those lines
of sight.  Given the large range of model curves, all measurements 
are consistent with some superposition of spinning dust, vibrational
dust, and free-free emission. 
}
\label{fig_result}
\end{figure*}

\end{document}